\documentclass[aps,prb,reprint,superscriptaddress]{revtex4-2}
\usepackage{amssymb,amsmath,bm,braket,txfonts,graphicx,enumerate,mathtools}
\usepackage[justification=Justified, format=plain, font=small]{caption}
\usepackage[justification=Justified, format=plain, font=small]{subcaption}
\usepackage[bookmarks=true,colorlinks,citecolor=blue,urlcolor=blue]{hyperref}
\usepackage{pgfplots}
\usepackage{tikz}
\usepgfplotslibrary{colorbrewer}
\usetikzlibrary{matrix} % <- added for self made legend
\pgfplotsset{compat=1.18}

\begin{document}
\title{Localized Fock Space Cages in Kinetically Constrained Models}

\author{Cheryne Jonay}
\affiliation{Department of Physics, Faculty of Mathematics and Physics, University of Ljubljana,
Jadranska 21, Ljubljana SI-1000, Slovenia}
\author{Frank Pollmann}
\affiliation{
Technical University of Munich, 
TUM School of Natural Sciences, 
Physics Department,
James-Franck-Str. 1,
85748 Garching,
Germany}
\affiliation{Munich Center for Quantum Science and Technology (MCQST), Schellingstr. 4, 80799 M{\"u}nchen, Germany}

\begin{abstract}
We investigate a mechanism for non-ergodic behavior in many-body quantum systems arising from destructive interference, leading to localization in Fock space. 
Drawing parallels with single-particle flat-band localization and Aharonov-Bohm cages, we identify conditions under which similar interference effects in the many-body domain produce \emph{Fock space cages} (FSCs)—highly localized many-body eigenstates. 
By interpreting Fock space as a graph where nodes represent bitstring basis states and edges denote non-zero transition amplitudes of the Hamiltonian, we analyze different kinetically constrained models. 
The FSCs cause non-ergodic dynamics when the system is initialized within their support, highlighting a universal interference-driven localization mechanism in many-body systems. 
\end{abstract}

\maketitle

The question of whether and how closed quantum many-body systems thermalize, typically described by the Eigenstate Thermalization Hypothesis (ETH)~\cite{Deutsch91, Srednicki94, Rigol2008, ETHreviewRigol16, Kim2014}, has obtained significant attention over the past years.
While generic quantum systems typically reach thermal equilibrium for local observables, this scenario breaks down in certain cases.
Notable examples include integrable systems~\cite{Rigol2007, Kinoshita2006} and many-body localized (MBL) phases~\cite{Basko06, Nandkishore14, AltmanVosk, Schreiber2015}.
In both cases, thermalization is circumvented due to an extensive set of conserved quantities~\cite{Essler16, Huse14, Serbyn13cons}.
These conservation laws lead in turn to non-ergodic dynamics, even at high energy densities.
A key question that has recently drawn significant attention concerns systems exhibiting intermediate behavior—neither fully localized nor completely ergodic. 
In this context, several novel classes of models displaying weak ergodicity breaking have reignited interest in quantum thermalization. 
These models are characterized by dynamics that depend strongly on initial conditions, distinguishing them from both fully thermal and strongly ergodicity-breaking systems.
Quantum many-body scars~\cite{Moudgalya01, Moudgalya02, ShiraishiMori, TurnerNatPhys, TurnerPRB, Choi2018}, arising from a distinct subset of non-thermal eigenstates embedded within a thermal spectrum, are a prominent example, observed in Rydberg atom experiments~\cite{Bernien2017, Bluvstein2021}.
Another example involves fractonic systems, such as one-dimensional (1D) models with conserved U(1) charge and associated dipole moment~\cite{Sala2020, Khemani2020, Moudgalya2019}, where the Hilbert space fragments into exponentially many disconnected subspaces---a phenomenon known as Hilbert-space fragmentation.

\begin{figure}[]
\centering
\includegraphics[width=\columnwidth]{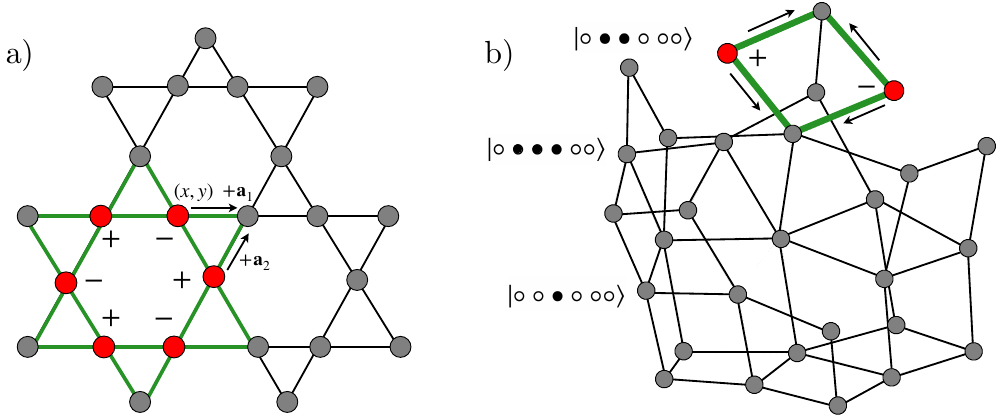}
\caption{(a) Single particle on the kagome lattice: destructive interference of nearest‐neighbor hoppings around a hexagon suppresses all amplitudes outside the green loop, localizing the single‐particle wavefunction. Red sites are charged, gray ones are neutral. (b) Many-body Fock space: a similar cancellation of transition amplitudes around a a closed loop in the many-body state space produces a \emph{Fock-space cage} (FSC). Here, nodes represent bitstring configurations, and edges represent non-zero matrix elements of the Hamiltonian.}\label{fig:Intro} %\fp{ To be consistent with (a), should we adjust the colors in (b)? Also we might want to adjust the arrows for this example?}
\end{figure}

In this work we investigate a mechanism that yields non-ergodic behavior due to destructive many-body interference effects, which in turn lead to \emph{exact eigenstates that are localized in Fock space}. 
This mechanism is akin to well-known phenomena in single-particle physics, such as localization in flat-band systems where destructive localizes particles.
Notably, flat bands emerging from lattice geometries like the kagome lattice (see Fig.~\ref{fig:Intro}a) or, more generally, from line graph constructions exhibit dispersionless energy bands, implying an extensive degeneracy of localized eigenstates \cite{Mielke1991,Tasaki1992,Leykam2018}. 
In these systems, the kinetic energy is quenched, and interference effects dominate the dynamics. 
Another prominent example are Aharonov-Bohm cages, where an applied magnetic flux can lead to complete localization due to destructive interference on specific lattice structures \cite{Vidal_1998,Mosseri_2022,Kang2020}.
We extend this concept to the many-body domain, demonstrating that many-body \emph{Fock space cages} (FSCs) can emerge, leading to localized many-body eigenstates.

To this end, we develop a graph‐theoretic framework for generic kinetically constrained models with chiral (particle–hole) symmetry. 
In Sec.~\ref{sec:GraphTheory}, we show how this symmetry endows the Fock‐space graph with a bipartite structure that lower‐bounds the number of zero‐energy modes via sublattice imbalance, and recast the search for these modes as a parity-check problem, which can be conducted either globally (at exponential cost) or locally (for non-exhaustive result).
In Sec.~\ref{sec:Models}, we illustrate this fraemwork in three types of kinetically constrained models, which differ in the number and locality of the cages they host.
Certain states are fully localized on $\mathcal{O}(1)$ nodes of the exponentially large Fock space graph and others span $\mathcal{O}(L)$ nodes.
Although the Hamiltonians remain globally ergodic—with typical (thermal) eigenstates spreading over an exponentially large fraction of Fock space—each model supports a set of localized zero‐energy cage states.
We show that the existence of such states leads to non-ergodic dynamics when the system is initialized in states with finite overlap with the FSCs.
While our detailed case studies focus on these three models, the same graph‐theoretic framework applies to a broader class of dynamically constrained models exhibiting an effective chiral or particle-hole symmetry, such as the East model \cite{Garrahan2009,Pancotti2020,Menzler2025}, the East-West model \cite{Brighi2024} and the PXP model \cite{TurnerPRB,Schecter2018,Buijsman2022}

\section{Graph-theoretical framework of FSCs}\label{sec:GraphTheory}
Before exploring specific models, we outline a general approach to identify FSCs on bipartite graphs using graph theory.
The Fock space naturally lends itself to a graph-based interpretation (see Fig.~\ref{fig:Intro}b): bitstring states correspond to nodes $v_n$, while the non-zero matrix elements of the many-body Hamiltonian $H_{nm}$ define the edges.
Disregarding the amplitudes of $H_{nm}$, this yields the connectivity structure of an undirected graph, where each vertex has degree $\mathcal{O}(L)$ for local Hamiltonians.
For kinetically constrained models with chiral or particle-hole symmetry, this graph is bipartite, splitting into even-parity (subalttice $A$) and odd-parity (sublattice $B$) states. This will help us bound the number of zero states, and also find some of them explicitly. 

\subsection{Number of zero states}
Hamiltonians with a chiral symmetry correspond to bipartite graphs, which in a suitable basis (organized by parity number) can be written as 
\begin{align}
    H\rightarrow \begin{pmatrix}
        0&M\\
        M^{\dagger}&0
    \end{pmatrix},
\end{align}
where $M$ is the biadjacency matrix linking $A$ to $B$, with dimensions $|A|$ and $|B|$ respectively. The number of zero-energy states is tied to the kernel of $M$ and $M^{\dagger}$ through the conditions $M \psi_B=0 = M^{\dagger}\psi_A$. By rank-nullity,
\begin{align}
\dim\ker H =\dim\ker M + \dim\ker M^\dagger \ge \bigl|\,|A| - |B|\,\bigr|,
\end{align}
providing a minimal number of zero-energy states determined by the sublattice imbalance. 
This principle applies broadly to systems with chiral symmetry, where the interplay between the symmetries can lead to an exponential number of zero-energy state, as observed in kinetically constrained models. 
For example, Ref.~\cite{Buijsman2022} derived exact lower bounds on zero-energy eigenstates using Fibonacci numbers for the PXP model \cite{TurnerNatPhys, TurnerPRB}, exploiting the combination of chiral and translation or inversion symmetry of the model. 
While these bounds are specific to the PXP model’s Hamiltonian and boundary conditions, the use of sublattice imbalance to count zero-energy states generalizes across systems with similar symmetries.

The bipartite nature of the Fock‐space graph has another useful consequence: if \(N_A \neq N_B\), any zero‐energy eigenstate must lie entirely on the larger sublattice.  We define the sublattice‐parity operator $\Gamma =\rm diag(
\mathbb{I}_{N_A},-\mathbb{I}_{N_B})$, which anticommutes with $H$, $\{\Gamma,H\}=\Gamma H + H \Gamma = 0.$
Hence, for any $\ket{\psi}$ satisfying $H\ket{\psi}=0$,
$H\bigl(\Gamma\ket{\psi}\bigr)
= -\,\Gamma\,H\ket{\psi}
= 0$, $\Gamma$ preserves the zero‐mode subspace and diagonalizes it: its $+1 (–1)$ eigenspace sits entirely on $A (B)$. We rely heavily on this property to find the FSC's algorithmically. 

\subsection{Identifying FSCs}\label{sec:Identifying_FSC}
To construct the Fock‐space cages explicitly, we exploit two facts: (i) zero‐modes are localized on either sublattice, and (ii) any nontrivial zero‐mode must form a closed loop (cage) in the bipartite graph.

Fact (i) allows us to recast the search for zero energy states into a simple backtracking algorithm. In principle, the matrix elements of the biadjacency matrix $M_{ba}$ encode the weights of the Hamiltonian, but for simplicity we treat it as a binary connectivity matrix: $M_{ba}=1$ if the two nodes are connected, and $0$ otherwise. Keeping the $B$ nodes set to $0$, and allowing the $A$ nodes to take on values $\{-1,0,+1\}$, a candidate zero mode on the A sublattice is a vector $\psi_A \in\{+1,0,-1\}^{|A|}$ such that each $B$ row sums to zero. By enforcing this zero‐sum constraint at every $B$-site, we automatically require that any nontrivial assignment closes into loops on the bipartite graph (fact ii). We initialize all $A$-nodes to be unassigned, then proceed column by column: at step $k$ we assign $(\psi_A)_k \in \{-1,0,1\}$ and immediately check the sum of every $B$-row whose A-neighbors have all been assigned. If any such row sum $\neq0$, we prune this branch immediately. If no row fails, we move to $k+1$. When $k=|A|$ and all row sums are satisfied, we have found a zero mode. We then backtrack to test for remaining assignments. Such backtracking with forward checking is a standard technique in the constraint-satisfaction and SAT-solving communities \cite{garey1979computers}, and it closely parallels decoding strategies for parity-check codes (e.g.\ LDPC decoding) \cite{gallager1962ldpc,mackay2003information}, although in our case the node assignments take values in a ternary set rather than being strictly binary. While the worst-case cost is exponential in $|A|$, the cages found for smaller systems often generalize to larger ones. 

When our goal is to identify primarily local cages rather than exhaustively enumerate every zero‐energy state, we devise a local ``charge‐flow'' algorithm that is far more efficient. We begin by fixing all $B$‐sites to zero and selecting a random seed $B$‐site (this choice can be refined by analyzing the connectivity of the graph). Next, we pick two of its neighboring $A$‐sites and assign them opposite charges, $+1$ and $–1$ (leaving all other $A$‐sites at $0$). This preserves neutrality at the seed but creates imbalances on its next nearest neighbor $B$‐sites. We continue the construction by iteratively neutralizing the $B$ nodes. For each charged $B$‐site, we assign compensating $\pm 1$ charges to one or more of its unassigned $A$‐neighbors until it is neutral. If a $B$‐site cannot be neutralized because no suitable $A$‐neighbors remain, we either backtrack to the previous assignment or restart with a new seed.

\begin{figure}[]
\centering
\includegraphics[width=\columnwidth]{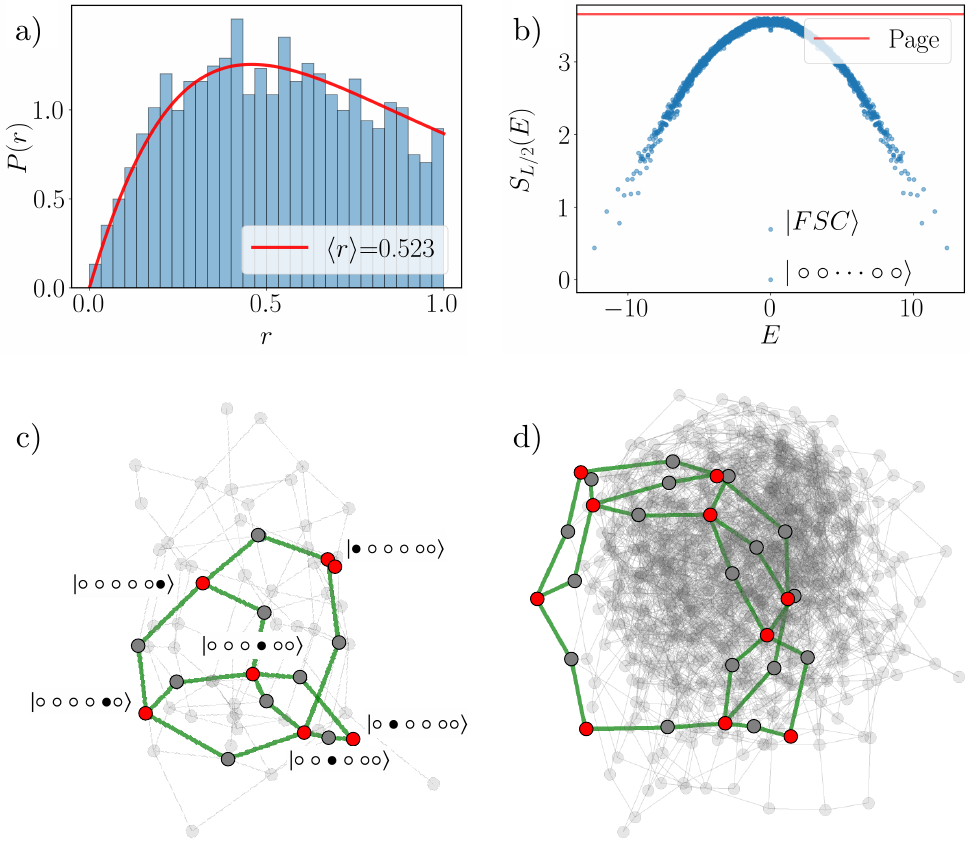}
\caption{(a) Distribution of the adjacent-gap ratio and (b) half-system entanglement entropy for the single-cage model Eq.~\eqref{eq:EWsingleCage} on $L=12$  with open boundary conditions. The trivial state $\ket{\circ\cdots\circ}$ (which spans its own Krylov subspace) and the FSC appear as outliers at zero energy. Despite the model’s global ergodicity, the cage shows sub-volume-law entanglement and remains localized in Fock space. (c)-(d) Visualization of the Fock-space cage Eq.~\eqref{eq:EWsingleCage} on the bipartite state-space graph for system sizes with open boundary conditions $L=6$ (c) and $L=10$ (d). The single-FSC lives on the red $A$ nodes, the blue $B$-nodes are neutral.}
\label{fig:EmW_2EmW}
\end{figure}

\section{FSCs in Kinetically Constrained models}\label{sec:Models}
We study three different models to highlight distinct aspects of FSCs. First, we consider a model that supports a single FSC across $\mathcal{O}(L)$ Fock states within an otherwise ergodic system. We then introduce two additional models that exhibit extensively many exact FSCs localized on $\mathcal{O}(L)$ and $\mathcal{O}(1)$ nodes, respectively.

\subsection{Model with single FSC of size $\mathcal{O}(L)$}
To illustrate the concept of FSCs, we begin with an extended ``East–West model''~\footnote{Note that our East-West model differs from the one introduced in Ref.~\cite{Brighi2024} in that it does not have a $U(1)$ symmetry but instead involves spin flips as in the original East model.}, which, as we will demonstrate, contains only a single cage and exhibits weakly ergodic behavior.
This makes it a minimal yet insightful setting for exploring the interplay between ergodicity and localization in dynamically constrained systems.
The model consists of a one-dimensional lattice with spin degrees of freedom, $\ket{0} = |\circ\rangle$ and $\ket{1} = \ket{\bullet}$ denoting empty and occupied sites, respectively.
Spin flips $X_i$ at site $i$ are governed by constraints involving either the nearest or next-nearest neighbors to the left (right).
The Hamiltonian reads
\begin{align}
H = J\sum_{i} \left[X_i(P_{i-1} - P_{i+1}) + X_i(P_{i-2} - P_{i+2})\right]
, \label{eq:EWsingleCage}
\end{align}
where $P_i = \frac{1 - Z_i}{2}$ is a projector onto the $\ket{\bullet}$ state.
All terms in the Hamiltonian anticommute with $\mathcal{P}= \prod_i Z_i$, and thus the model exhibits a chiral symmetry.
The choice of boundary conditions plays an important role: to access the single-cage regime, translation invariance must be broken, which we achieve by imposing open boundary conditions.
The model transitions from a multi-cage to a single-cage structure at zero energy either by breaking translation symmetry (e.g., via boundaries) or by introducing disorder.
Note the combination of a chiral symmetry together with translation or inversion symmetry leads to an exponential number of zero energy states as discussed above.
\begin{figure}[]
\centering
\includegraphics[width=\columnwidth]{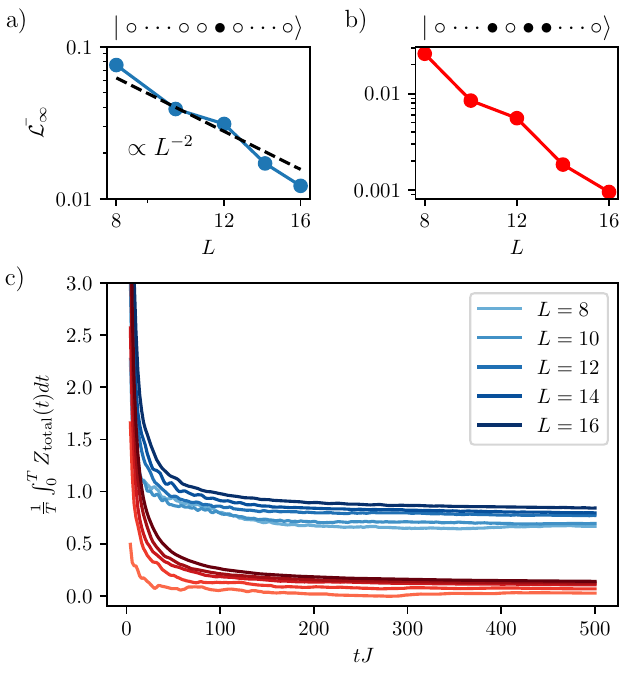}
\caption{Quench dynamics of the single cage model Hamiltonian Eq.~(\ref{eq:EWsingleCage}) with open boundary conditions, starting from two different initial states for various system sizes $L$. (a) Saturation value of the return probability for a bit string that has order $\frac{1}{\sqrt{L}}$ overlap with the FSC state (blue) and (b) for a bit string that has no overlap with it (red). The dashed line in (a) shows the expected slope of $\propto L^{-2}$ as a guide to the eye. (c) Time-averaged expectation value of the magnetization is shown for the two states.}
\label{fig:EmW_2EmW_dynamics}
\end{figure}

Despite the strong kinetic constraints of the model, the system does not exhibit Hilbert space fragmentation \cite{Sala2020,Khemani2020,Moudgalya2019}.
In fact, apart from the trivial empty state, all remaining $2^L - 1$ configurations lie in a single connected Krylov sector under $H$.
We therefore restrict to the $(2^L - 1)$-dimensional Krylov subspace generated from the seed state $\ket{s_0}=\ket{\bullet\circ\cdots\circ}$.
Moreover, the Hamiltonian is chaotic and exhibits level-spacing statistics consistent with random-matrix theory.
The average adjacent-gap ratio is defined as $r_n = \frac{{\rm min}(\Delta E_n,\Delta E_{n+1})}{{\rm max}(\Delta E_n,\Delta E_{n+1})}$,
where $\Delta E_n = E_n-E_{n-1}$ \cite{Oganesyan_2007,Atas_2013}, and yields $\langle r\rangle \approx 0.53$, as shown in Fig.~\ref{fig:EmW_2EmW}a.
This is indicative of the Gaussian orthogonal ensemble (GOE); chiral symmetry suggests a slight shift toward chGOE, but numerically $\langle r\rangle$ remains near 0.53.
Next, we consider the entanglement of eigenstate $\psi_n$ at energy $E_n$, quantified by the von Neumann entropy of its reduced density matrix on a half chain.
As seen in Fig.~\ref{fig:EmW_2EmW}b, the entanglement entropy follows the Page curve \cite{Page1993} for mid-spectrum states, with two clear outliers: the empty state ($S=0$), which lies in its own Krylov sector, and the non-thermal FSC.

While the Krylov space scales exponentially with $L$, the zero-energy FSC is localized on $L$ nodes:
\begin{align}
    \ket{\text{FSC}} & = \sum_{i=0}^{L-1} \ket{2^i}  = \frac{1}{\sqrt{L}}\left(\ket{\bullet \circ \dots \circ}  + \cdots +   \ket{\circ \circ \dots \bullet}\right).\label{eq:FSC}
\end{align}
Here we introduce a binary representation of the bit strings such that $|2^i\rangle$ corresponds to bitstrings with a single 1 at position $i$.
The FSC behaves as a simple $k=0$ single-particle state—despite the Hamiltonian not conserving particle number.
Its exact eigenstate nature arises from perfect destructive interference, which dynamically confines the system to this localized cage state of $L$ nodes.
Figures~\ref{fig:EmW_2EmW}c and \ref{fig:EmW_2EmW}d illustrate this localization for $L=6$ and $L=10$, respectively. 
Despite residing in the middle of the spectrum, the FSC exhibits area-law entanglement $S=\log2$ well below the Page value as shown in Fig.~\ref{fig:EmW_2EmW}b.
The FSC state thus represents an exact scar state.

Next we study the effect of FSC on the real‐time dynamics.
We begin by examining the return probability  
\begin{align}
    \mathcal{L}(t) = |\langle\psi(0)|\psi(t)\rangle|^2
    , \label{eq:return}
\end{align}
starting from a simple bit-string configuration \( |\psi(0)\rangle = |\circ \cdots\ \circ\ \bullet\ \circ\ \cdots\ \circ\rangle \). This initial state has an overlap of \( 1/\sqrt{L} \) with the \( |\text{FSC}\rangle \), leading us to expect that  
$|\langle\psi(0)|\psi(t \rightarrow \infty)\rangle|^2 \propto \frac{1}{L^2}$,
in contrast to the exponentially small values characteristic of ergodic systems.
This expectation is confirmed by the numerical results presented in Fig.~\ref{fig:EmW_2EmW_dynamics}a.
Next, we analyze the time evolution of the total magnetization,  
\begin{align}
    Z_{\text{total}} = \sum_{i=1}^L \langle \psi(t) |\, Z_i \,| \psi(t) \rangle, \label{eq:mag}
\end{align}
where \( Z_i \) denotes the Pauli \( Z \) operator at site \( i \). 
As before, we use the same bit-string configuration as the initial state.  
Figure~\ref{fig:EmW_2EmW_dynamics}b displays the behavior of \( Z_{\text{total}}(t) \), which saturates to a value of order one rather than decaying to zero, as would be expected in an ergodic system without conservation of \( Z_{\text{total}} \).  
This behavior can once again be attributed to the non-negligible overlap between the initial and the FSC state.

\subsection{Model with multiple FSCs of size $\mathcal{O}(L)$}
\begin{figure}[]
\centering
\includegraphics[width=\columnwidth]{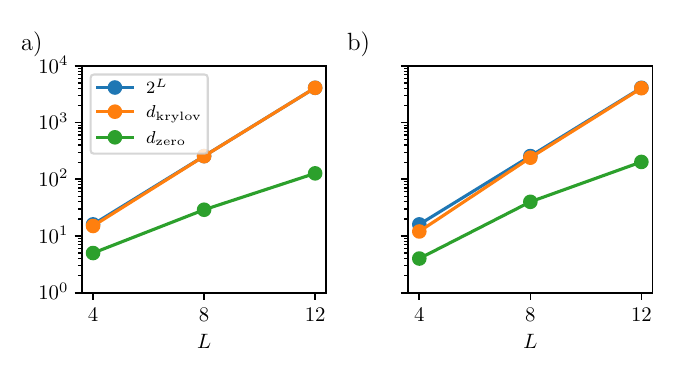}
\caption{Scaling of the Krylov subspace $d_{\text{krylov}}$ dimension  and the number of zero energy eigenstates $d_{\text{zero}}$ for Hamiltonians (a) Eq.~(\ref{eq:EWMultiCage}) and (b) Eq.~(\ref{eq:O1cage}) with periodic boundary conditions.}
\label{fig:Multi_cage_num_zeros}
\end{figure}
We now discuss a model that exhibits an extensive number of $\mathcal{O}(L)$ FSC states, described by the Hamiltonian 
\begin{align}
H = J \sum_i  X_i \, (P_{i-1} + P_{i+1}). \label{eq:EWMultiCage}
\end{align}
As in the previously considered model, all terms in the Hamiltonian anti-commute with $\mathcal{P}= \prod_i Z_i$, and thus the model exhibits chiral symmetry.
In addition to translation symmetry, this model also has inversion symmetry, which protects an extensive number of zero states even in the presence of open boundary conditions.
The scaling of the number of zero states is shown in Fig.~\ref{fig:Multi_cage_num_zeros}a and shows a clear exponential growth with system size. 
Apart from the trivial empty state, all remaining $2^L - 1$ configurations lie in a single connected Krylov sector under $H$.
We have verified that the level statistics within momentum and inversion sectors follow the GOE.
The dynamics (not shown) behaves similar to the single cage model in that states which have an overlap with any of the FSC states show non-ergodic behavior while other states appear to thermalize.

\begin{figure}[]
\centering
\includegraphics[width=\columnwidth]{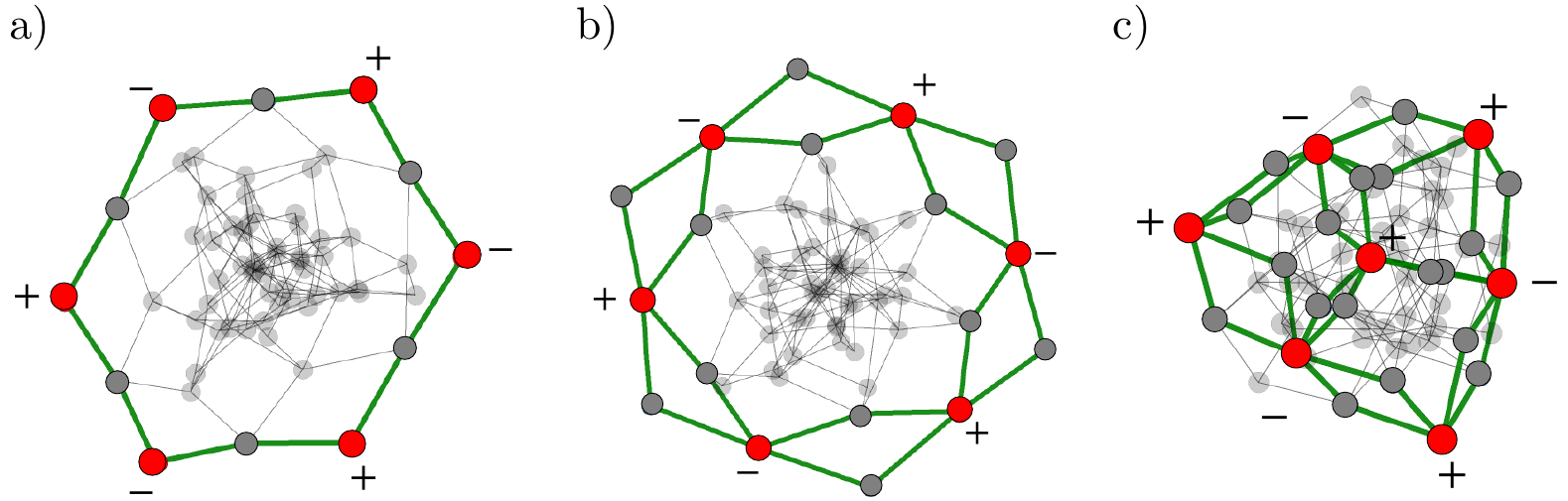}
\caption{FSC of size $\mathcal{O}(L)$ for the multi‐cage model Eq.~\eqref{eq:EWMultiCage} for $L=6$ spins on the $2^L-1$ dimensional Fock space graph: (a) $\ket{\text{FSC}_1}$, (b) $\ket{\text{FSC}_2}$, and (c) $\ket{\text{FSC}_3}$. Each panel displays a distinct cage on the bipartite graph. These cages grow in size with L and are present for all system sizes. Nodes with zero amplitude (i.e., outside the cage) are shown in gray.}%\fp{We should use colors that indicate the sign structure of the $k=\pi$ state.}}
\label{fig:OLCage}
\end{figure}

We can use the algorithm outlined in \ref{sec:Identifying_FSC} and explicitly construct the FSCs. 
Below, we list some closed‐form expressions for FSC’s of the model using the binary representation of bitstrings:
\begin{align}
    \ket{\rm FSC_1} &= \sum_{i=0}^{L-1} (-1)^i\ket{2^i}\\
    \ket{\rm FSC_2} &= \sum_{i=0}^{L-1}(-1)^i \ket{2^i+2^{(i+1)\bmod L}} \\
   %\ket{S_3} &= \sum_{i=0}^{L-1} (-1)^{\frac{i(i-1)}{2}} \left(|2^i\rangle + |(2^L-1)-2^i\rangle\right).\\
   \ket{\rm FSC_3} &= \ket{2^L-1} + \sum_{i=0}^{\tfrac L2 -1}\ket{2^i + 2^{\,i+L/2}} \\
      &\quad-\sum_{i=0}^{\tfrac L2 -1}\ket{(2^L-1)-\bigl(2^i + 2^{\,i+L/2}\bigr)}\notag 
\end{align}

$\ket{\text{FSC}_2}$ and $\ket{\text{FSC}_3}$ provide two linearly independent solutions each (by alternating the sign pattern). 
All the states identified have a support of $\mathcal{O}(L)$ nodes and are shown in Fig.~\ref{fig:OLCage}.

\subsection{Model with multiple FSCs of size $\mathcal{O}(1)$}

We now discuss a model that exhibits an extensive number of $\mathcal{O}(1)$ FSC states, described by the Hamiltonian

\begin{align}
H &= J\sum_{i\ \text{even}} X_{i}(P_{i-1} + P_{i+1}) \nonumber \\
  &\quad + J\sum_{i} X_{i}(P_{i-1}P_{i-2}P_{i-3} + P_{i+1}P_{i+2}P_{i+3}).
 \label{eq:O1cage}
\end{align}
Note that the first term is the same as in the previous model except that it is summed only over even sites $i$.
Like the models considered previously, this model exhibits chiral, translation, and inversion symmetries, which protect an extensive number of zero modes, as demonstrated in Fig.~\ref{fig:Multi_cage_num_zeros}b.
Apart from the trivial empty state, there exist additional $ 2^{L/2}-1$(frozen) states that are not connected to the largest Krylov sector, but they still constitute a set of measure zero compared to the Hilbert space dimension of $2^L$. 
These states correspond to product states with $\circ$ on all odd sites and an arbitary number of $\bullet$ on the even sites.

One can now identify $\mathcal{O}(1)$ FSCs, which are of the form::
\begin{align}
    \ket{\text{FSC}} = \frac{1}{\sqrt{2}}\left(\ket{\circ \cdots \circ\bullet\bullet \circ\circ \dots \circ} - \ket{\circ \cdots \circ\circ\bullet\bullet\circ  \dots \circ} \right).\label{eq:FSCO1}
\end{align}
See also Fig.~\ref{fig:OLCage} for a visualization of the FSC.
In addition to this pattern, there can be an arbitrary number of isolated $\bullet$ on the even sublattice or other spatially separated cages, leading to an exponential number of such cage states in the spectrum.
Moreover, there exist many larger FSCs that we do not list here. As in the previous model, the level spacings (not shown) are ergodic within momentum and parity sectors. 

\begin{figure}[]
\centering
\includegraphics[width=\columnwidth]{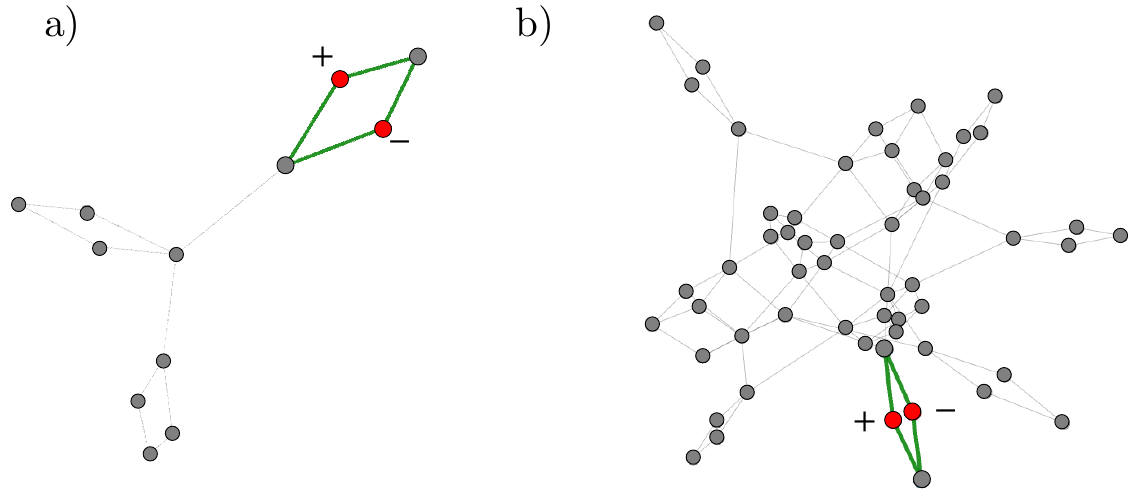}
\caption{FSC of size $\mathcal{O}(1)$ for Hamiltonian Eq.~\eqref{eq:O1cage} in the largest Krylov sector for system sizes (a) $L=4$ and (b) $L=6$. Nodes with zero amplitude (i.e., outside the cage) are shown in gray.}
\label{fig zo :O1cage}
\end{figure}

\begin{figure}[]
\centering

\includegraphics[width=\columnwidth]{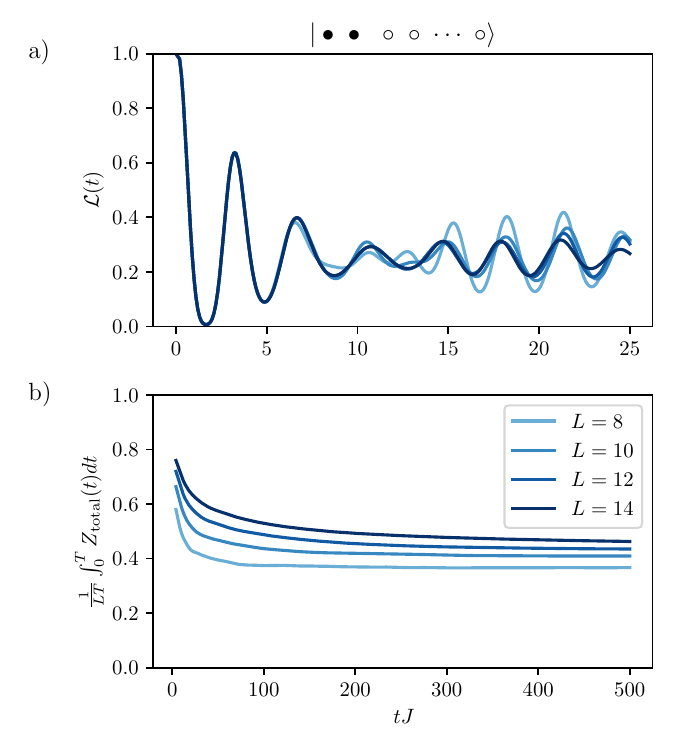}
\caption{Quench dynamics of the $\mathcal{O}(1)$ cage model starting from an initial state that has order one overlap with the $\mathcal{O}(1)$ FSC for Hamiltonian Eq.~(\ref{eq:O1cage}) for different system sizes $L$ with periodic boundary conditions. (a) Return probability and (b) time-averaged expectation value of the magnetization  $Z_{\text{total}}/L$.}
\label{fig:O1cage_dyn}
\end{figure}
Finally we explore  the effect of the $\mathcal{O}(1)$ cages on the dynamics.
Figure~\ref{fig:O1cage_dyn}a shows the return probability of the simple bit-string configuration $(\ket{\circ \cdots \circ\bullet\bullet \circ\circ \dots \circ}$. 
Since it has an overlap of \( 1/\sqrt{2} \) with  \( |\text{FSC}\rangle \), we obtain
$|\langle\psi(0)|\psi(t \rightarrow \infty)\rangle|^2 \approx \frac{1}{4}$,
in contrast to the exponentially small values characteristic of ergodic systems.
Figure~\ref{fig:EmW_2EmW_dynamics}b shows the evolution of the magnetization \( Z_{\text{total}}(t)/L \), which saturates to a value of order one rather than decaying to zero.
Both results are a direct consequence of the strong ($\mathcal{O}(1)$) localization of the FSC.

\section{Conclusion}
In this work, we have investigated a mechanism that leads to non-ergodic behavior due to destructive many-body interference effects, which in turn lead to localized states in Fock space.  
For this we developed a graph-theoretic framework for uncovering Fock-space cages (FSCs) - exact, localized, zero-energy eigenstates - in dynamically constrained many-body systems. 
By exploiting the bipartite structure of the Fock space graph induced by the chiral symmetry of these models, we can lower bound the number of zero energy states, as well as use algorithms to find them. 
We introduce two algorithms, one that does an extensive search, and a more efficient, local search. 
We applied these methods to three representative models: a single cage model supporting one $\mathcal{O}(L)$ FSC within an otherwise ergodic spectrum, and two multi-cage models that each host exponentially many FSCs, but differ in whether those are $\mathcal{O}(L)$ or $\mathcal{O}(1)$ localized. 
In every case, these cages imprint clear non-ergodic signatures in the dynamics - most notably in the long-time return probability and steady-state magnetization---even though the Hamiltonian as a whole remains ergodic.

Our results reveal a universal interference-driven localization mechanism in Fock space, closely analogous to flat-band and Aharonov–Bohm cages in single-particle physics, yet acting in the exponentially large many-body Hilbert space. 
Although we focused on localized FSC at zero energy in the paper, they also occur at finite energies.
Looking forward, a number of open questions remain: a systematic spectral-graph analysis of cage subgraphs, a complete classification of all FSCs in broader model classes, and the inverse problem of engineering Hamiltonians with prescribed cages. 
A promising setup for realizing FSC experimentally are Rydberg platform in which constrained dynamics can be implemented naturally.

\section{Acknowledgements} 
We thank Tom Ben-Ami, Markus Heyl, Roderich Moessner, and Alexander Altland for insightful discussions. 
CJ acknowledges Shashwat Silas for discussions highligthing the connection to the parity-check algorithm.
CJ acknowledges support from the European Union’s HORIZON-CL4-2022-QUANTUM-02-SGA program under the PASQuanS2.1 project (Grant Agreement No. 101113690). 
CJ and FP  acknowledge support from the Deutsche Forschungsgemeinschaft (DFG) through FOR 5522 (Project-ID 499180199). 
F.P.~acknowledges the DFG Germany’s Excellence Strategy--EXC--2111--390814868, TRR 360, as well as the Munich Quantum Valley, which is supported by the Bavarian state government with funds from the Hightech Agenda Bayern Plus.

\emph{Note added:} While preparing the manuscript, we
became aware of related works by Tao-Lin
Tan, and Yi-Ping Huang \cite{tan2025}, Tom Ben-Ami, Markus Heyl, and Roderich Moessner \cite{benami2025many}, and Eloi Nicolau, Marko Ljubotina, and Maksym Serbyn \cite{nicolau2025}.

\let\oldaddcontentsline\addcontentsline
\renewcommand{\addcontentsline}[3]{}
\bibliographystyle{apsrev4-1}
\bibliography{bib}
\let\addcontentsline\oldaddcontentsline

\newpage
\appendix

\end{document}